\pdfoutput=1

\documentclass{article}
\usepackage{ismir,amsmath,cite}
\usepackage{graphicx}
\usepackage{color}
\usepackage{tabularx}
\usepackage[table]{xcolor}
\usepackage{babel,blindtext}
\usepackage{atbegshi}
\usepackage{amssymb}

\usepackage{url}
\usepackage{tikz}
\usepackage[table]{xcolor}    

\usepackage{subfigure}

\usepackage{tikz}

\usepackage{multirow}
\usepackage{adjustbox}

\usepackage{lineno}

\makeatletter
\def\thickhline{%
  \noalign{\ifnum0=`}\fi\hrule \@height \thickarrayrulewidth \futurelet
   \reserved@a\@xthickhline}
\def\@xthickhline{\ifx\reserved@a\thickhline
               \vskip\doublerulesep
               \vskip-\thickarrayrulewidth
             \fi
      \ifnum0=`{\fi}}
\makeatother

\newlength{\thickarrayrulewidth}
\DeclareMathOperator*{\argmax}{argmax} 

\newcommand{\mc}[2]{\multicolumn{#1}{#2}}
\usepackage{booktabs,siunitx}
\tikzstyle{bag} = [align=center]

\title{MSTRE-Net: Multistreaming Acoustic Modeling for Automatic Lyrics Transcription}

\threeauthors
  {Emir Demirel} {Queen Mary University of London\\ {\tt e.demirel@qmul.ac.uk}}
  {Sven Ahlb\"ack} {Doremir Music Research AB\\ {\tt sven.ahlback@doremir.com}}
  {Simon Dixon} {Queen Mary University of London\\ {\tt s.e.dixon@qmul.ac.uk}
  \thanks{ED received funding from the European Union’s Horizon 2020 research
and innovation programme under the Marie Skłodowska-Curie grant agreement No. 765068.}}

\begin{document}

\maketitle
\begin{abstract}

This paper makes several contributions to automatic lyrics transcription (ALT) research. Our main contribution is a novel variant of the Multistreaming Time-Delay Neural Network (MTDNN) architecture, called MSTRE-Net, which processes the temporal information using multiple streams in parallel with varying resolutions keeping the network more compact, and thus with a faster inference and an improved recognition rate than having identical TDNN streams. In addition, two novel preprocessing steps prior to training the acoustic model are proposed. First, we suggest using recordings from both monophonic and polyphonic domains during training the acoustic model. Second, we tag monophonic and polyphonic recordings with distinct labels for discriminating non-vocal silence and music instances during alignment. Moreover, we present a new test set with a considerably larger size and a higher musical variability compared to the existing datasets used in ALT literature, while maintaining the gender balance of the singers. Our best performing model sets the state-of-the-art in lyrics transcription by a large margin. For reproducibility, we publicly share the identifiers to retrieve the data used in this paper.

\end{abstract}

\section{Introduction}\label{sec:introduction}


Empirical studies show that it is a challenging task even for human listeners to recognize sung words, and this is more challenging than speech, due to a number of performance, environment and listener related factors\cite{fine2014making}.  Thus the automatic retrieval of sung words through machine listening, i.e.\ automatic lyrics transcription (ALT), can potentially be impactful in easing some of the time consuming processes involved in composing music, audio/video/music score captioning and editing, lyrics alignment, music catalogue creation, etc. Despite its potential, the current state of lyrics transcription is far from being sufficiently robust to be leveraged in such applications.

With recent advances in automatic speech recognition (ASR) research and its successful adaptation to singing data, considerable improvements have been reported in ALT research \cite{dabike2019automatic,gupta2018semi,demirel2020automatic}. In addition to this, newly released datasets have accelerated the development of the research field\cite{meseguer2019dali,damp}. Through these improvements, the prospect of applying ALT in the music industry has become more realistic, assuming that progress continues. Although promising results have been obtained for a cappella recordings\cite{dabike2019automatic,demirel2020automatic,demirel2021pron}, recognition rates drop considerably in the presence of instrumental accompaniment\cite{demirel2021low,gupta2020automatic}.

From the perspective of ASR, music accompaniment can be regarded as noise since non-vocal music signals generally include minimal or no information relevant to lyrics transcription, while their presence in the spectral domain increases the confusions during prediction. For building more robust acoustic models against noisy environments, the multistream approach in ASR was introduced \cite{bourlard1996mew}, inspired by how the acoustic signals are split into multiple frequency bands and processed in parallel in the human auditory system\cite{allen1994}. While previous research suggested using multiresolution feature processing\cite{hermansky2005multi,tuske2018acoustic} or reconstruction of a multi-band latent representation through autoencoders\cite{mallidi2015autoencoder} to achieve multistreaming ASR, the neural network architecture recently introduced in \cite{han2019state}, Multistreaming Time-Delay Neural Network (MTDNN), proposes a simplified solution which is utilized in producing the state-of-the-art for hybrid / Deep Neural Network - Hidden Markov Model (DNN-HMM) based ASR\cite{han2020multistream,pan2020asapp}. In this work, we propose a compact variant of MTDNN, referred to as MSTRE-Net,  where streams are diversified by having different numbers of layers with the goal of reducing the number of trainable parameters (i.e. model complexity), and thus the inference times and improving the word recognition rates. 

Additionally, we propose a number of other novel contributions for improving lyrics transcription performance. We suggest combined training of the acoustic model on both monophonic (e.g.\ $\textit{DAMP}$-Sing! 300x30x2 \cite{damp}) and polyphonic (e.g.\ $\textit{DALI}$ \cite{meseguer2019dali}) recordings, which is shown to improve performance for both cases. Furthermore, we propose tagging monophonic and polyphonic utterances with separate \emph{music} and \emph{silence} tokens explicitly. Our goal for this is to generate alignments that are more robust against disruptions in the decoding path, potentially caused by the musical accompaniment during the non-vocal frames. 

One major challenge in ALT research has been publishing reproducible results, due to the lack of publicly available evaluation data\cite{kruspe2015training}. Dabike and Barker \cite{dabike2019automatic} shared manually verified annotations for a subset of $\textit{DAMP}$ which have been utilized for evaluation in a cappella singing\cite{demirel2020automatic,demirel2021pron}. The Jamendo (lyrics) dataset \cite{stoller2019end} consists of 20 contemporary polyphonic music recordings released under an open source license. Moreover, despite their limited nature in terms of size and musical variability, Hansen\cite{hansen2012recognition} and Mauch\cite{mauch2011integrating} have been among the two most commonly used evaluation sets for ALT. In addition to these, we present a new test set with 240 polyphonic recordings having a larger span of release dates and better singer gender balance in order to establish a more comprehensive lyrics transcription evaluation. 

The rest of the paper is structured as follows: we begin with a summary of essential concepts in the state-of-the-art approach for hybrid-ASR. The following section explains how the proposed MTDNN architecture is constructed. Next, we give details of the data used in experiments, and introduce a new evaluation set. Finally, we describe the experimental setup and present results verifying our design choices through ablative tests.


\section{Background}

ALT can be considered as analogous to Large Vocabulary Continuous Speech Recognition (LVCSR) for the singing voice. Similarly, the goal for ALT is predicting the most likely word sequence, $\mathbf{w}$, given a stream of acoustic observations, $\mathbf{O}$, which can be expressed in mathematical terms as follows:

\vspace{-.5cm}

 \begin{align}\label{eq:asr}
    \widehat{\mathbf{w}} =  &= \argmax_{\mathbf{w}} P(\mathbf{w}|\mathbf{O}) \nonumber \\
    &= \argmax_{\mathbf{w}} P(\mathbf{w}) p(\mathbf{O}|\mathbf{w}) \nonumber \\
                         &= \argmax_{\mathbf{w}} P(\mathbf{w}) \sum_\mathbf{Q} p(\mathbf{O}|\mathbf{Q}) P(\mathbf{Q}|\mathbf{w}), 
\end{align}

\vspace{-0.2cm}

\noindent where elements of $\mathbf{Q}$ represent the phoneme classes\footnote{A phoneme is the basic sonic unit of speech. In linguistics, words are considered to be composed of sequences of phonemes.}. In Equation \ref{eq:asr}, $p(\mathbf{O}|\mathbf{Q})$ is obtained via the acoustic model. Phonemes are converted to word labels using a lexicon which defines a mapping between words and their phonemic representations. The raw word posteriors are then smoothed with the language model, $P(\mathbf{w})$ for obtaining grammatically more plausible output transcriptions, $\widehat{\mathbf{w}}$.

According to the probabilistic approach of ASR, phonemes are represented with HMMs where a transition between connected phone states occurs at every time step\cite{gales2008application}. In our system, we employ the \textit{Kaldi} toolkit \cite{povey2011kaldi}, an open-source ASR framework that represents HMM states using Weighted Finite State Transducers (WFST)\cite{mohri2002weighted}. In operation, a  WFST graph is generated through composing posteriors retrieved from the acoustic, language and pronunciation models. The resulting directed paths of states form a \textit{lattice}, a weighted acyclic graphical structure which can represent multiple output hypotheses.

\begin{figure}[!t]
 \centering
 \includegraphics[clip,width=0.4\textwidth,height=4.3cm]{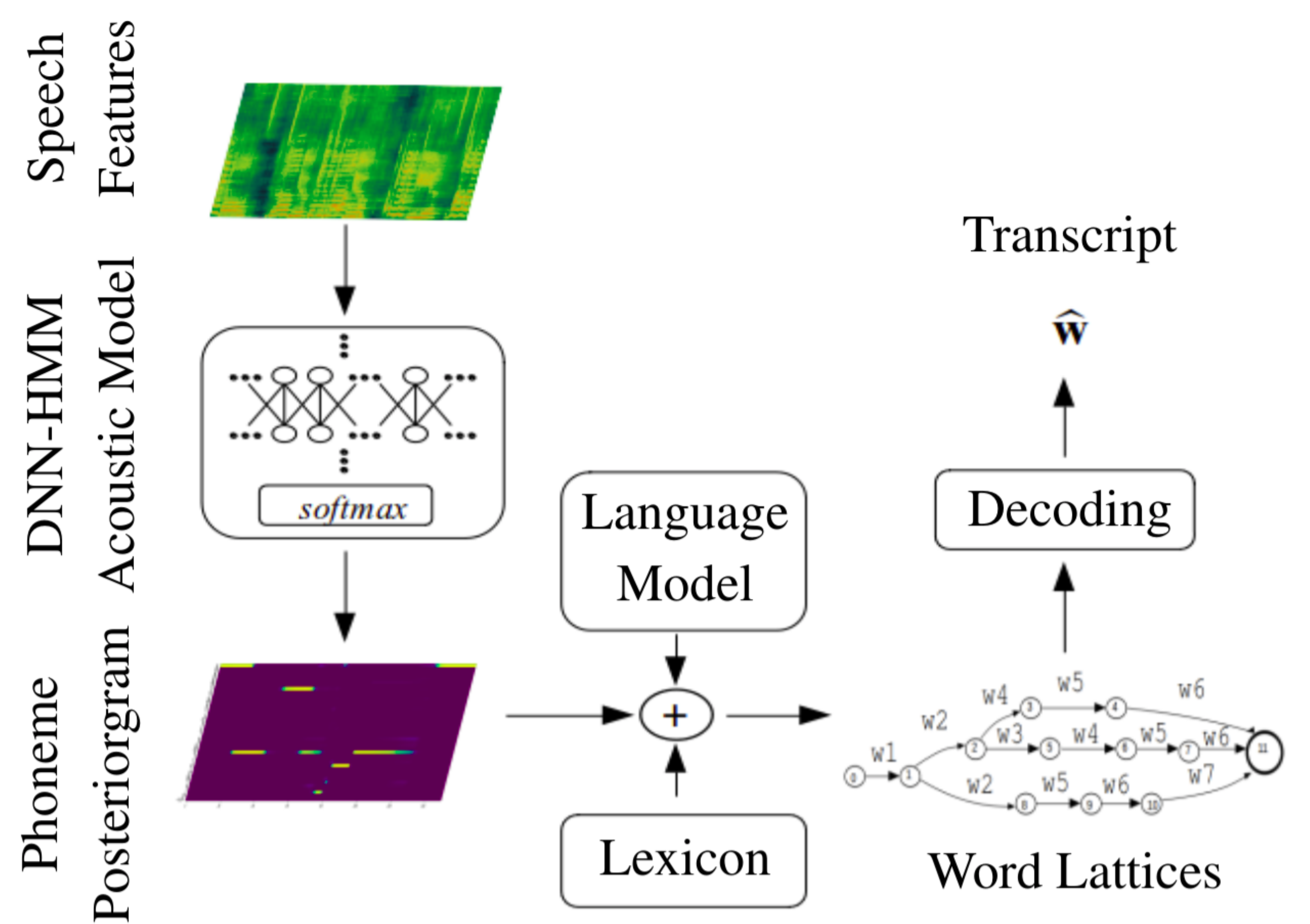}%
 \caption{DNN-HMM based ASR at operation}
 \label{fig:sys_text}
\end{figure}

\subsection{Lattice-free Maximum Mutual Information}\label{sec:lfmmi}

Most recent ALT systems utilize the state-of-the-art hybrid DNN-HMM framework where the neural networks are trained in a sequence discriminative fashion\cite{vesely2013sequence}. More specifically, the best performing lyrics transcribers to date \cite{dabike2019automatic,gupta2018semi,demirel2020automatic,demirel2021low,demirel2021pron} use Lattice-free Maximum Mutual Information training (LF-MMI)\cite{povey2016purely}, where network parameters are tuned w.r.t.\ the MMI objective: 

\vspace{-.15cm}

 \begin{equation}\label{eq:mmi}
    \mathcal{F}_{\textit{MMI}} = \sum_u  \log \frac{p(\mathbf{O}_u| Q_u)^\mathcal{K} P(W_u)}{\sum_W  p(\mathbf{O}_u|Q)^\mathcal{K} P(W)}
\end{equation}

\noindent where $p(\mathbf{O}_u| Q_u)$ is the probability of observing an acoustic instance $O$ in the utterance $u$, in Markovian phone state $Q_u$, and the $P(W)$'s are the word sequence probabilities\cite{bahl1986maximum}. Optimization w.r.t.\ MMI aims at maximizing the shared information between the reference and target sequences. More explicitly, the terms in the numerator are calculated per utterance, whereas the denominator is computed over the entire training set. Hence, the network parameters are updated to maximize the probabilities in the numerator and minimize the denominator. In other words, the goal of MMI training is to discriminate a certain acoustic observation with its given utterance.

\section{Multistreaming Time-Delay Neural Networks}\label{sec:mtdnn}

\begin{figure*}[!ht]
\centering
\subfigure[Single-stream TDNN ($\tau$=3)]{%
\label{fig:first}%
\includegraphics[height=3cm,width=0.3\textwidth]{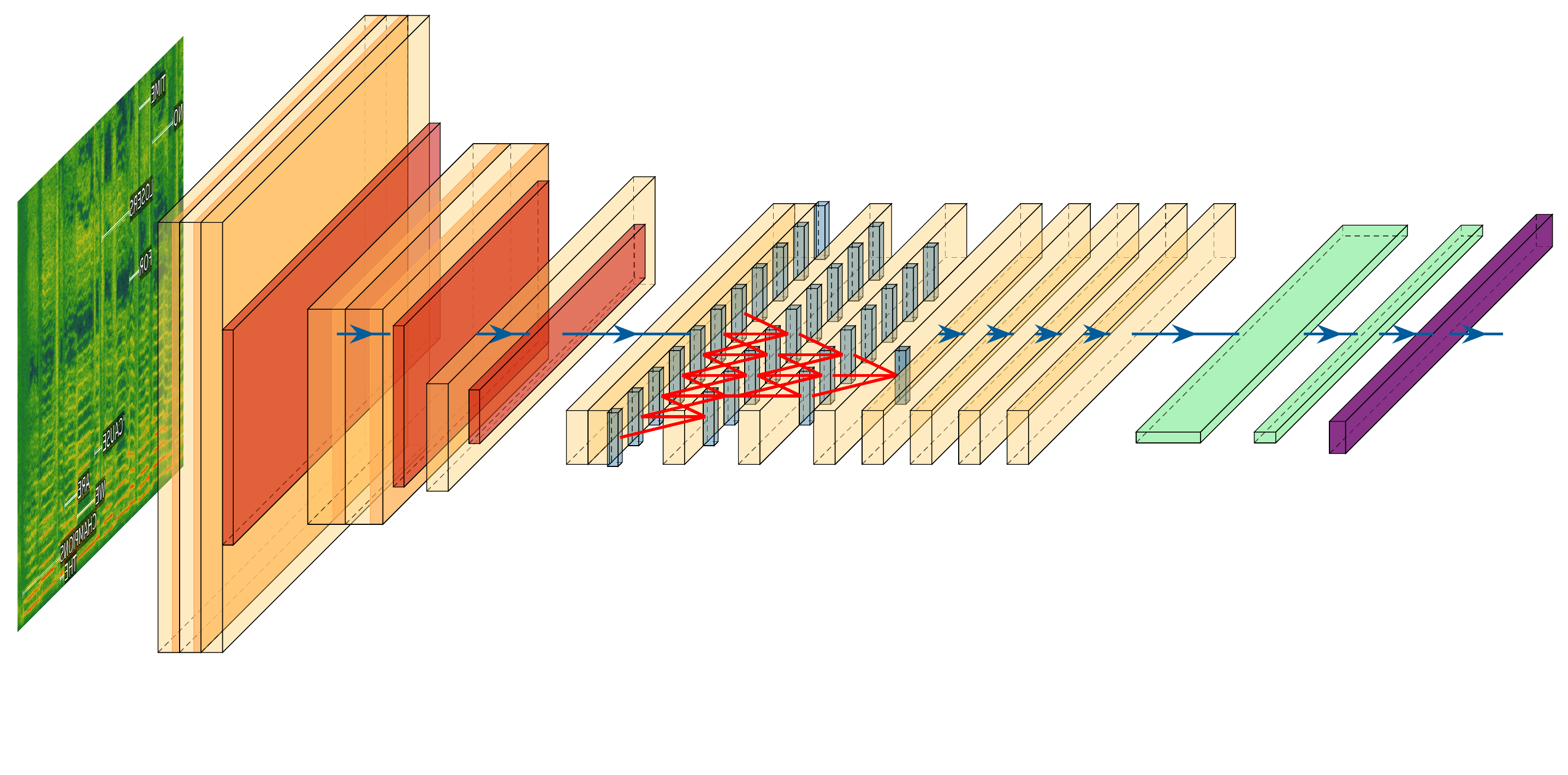}}
\subfigure[MTDNN with identical streams ($\tau$=\{3,6,9\})]{%
\label{fig:second}%
\includegraphics[height=4.4cm,width=0.35\textwidth]{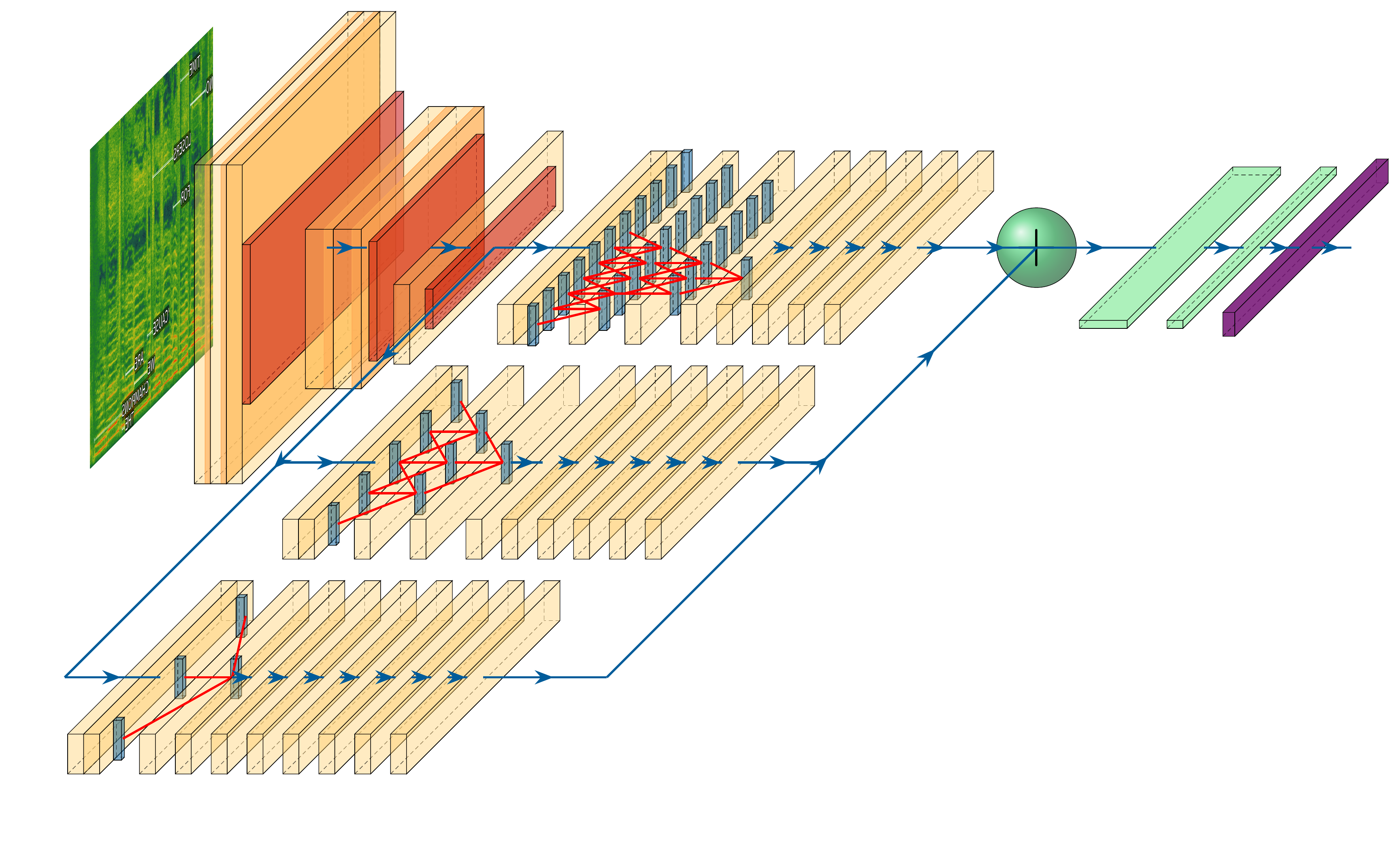}}
\subfigure[MTDNN with distinct streams ($\tau$=\{3,6,9\})]{%
\label{fig:third}%
\includegraphics[height=4.4cm,width=0.35\textwidth]{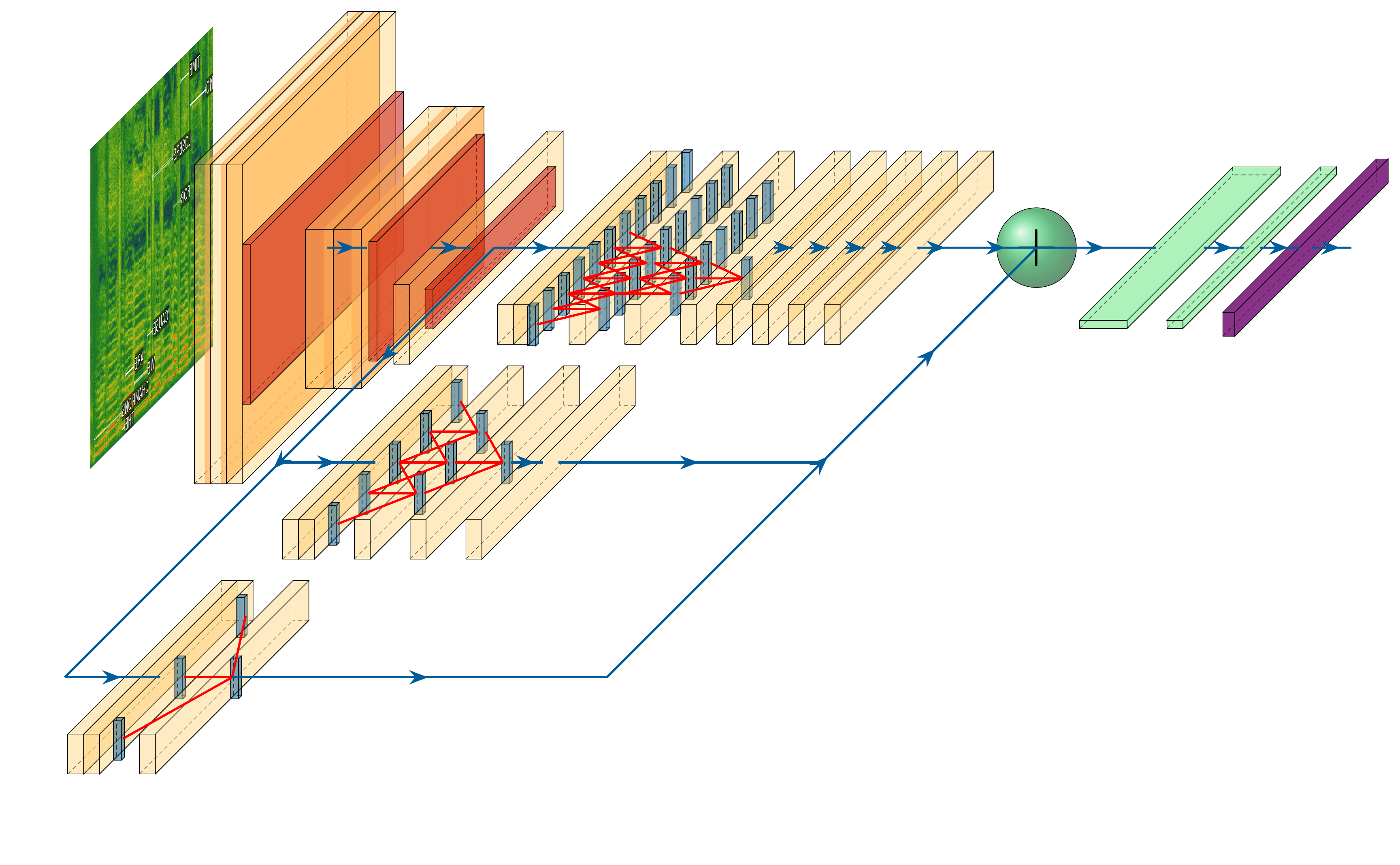}}%
\caption{Different variants of TDNN architectures. From left-to-right, the orange, yellow and cyan blocks represent the front-end 2-D CNN, TDNN streams and final FC layers preceded by the purple softmax layer} 
\label{fig:mtdnn}
\end{figure*}

The main body of MTDNN architectures consists of multiple streams of TDNN layers trained in parallel, where each stream has a unique time dilation rate ($\tau$). Our proposed MTDNN variant differs from the original models \cite{pan2020asapp,han2020multistream} by having different numbers of layers in the TDNN streams, depending on $\tau$ (Figures \ref{fig:second} and \ref{fig:third}).

Prior to the TDNN streams, input features $\mathbf{x}$ are first processed by a single stream 2-D Convolutional Neural Network (CNN) in the front-end of the network, 
\vspace{-.22cm}

 \begin{equation}
    \mathbf{h}= \mathit{Stacked{\text -}2D{\text -}CNN}(\mathbf{x})
\end{equation}

\noindent where $\mathit{Stacked{\text -}2D{\text -}CNN}$ stands for the stack of 2-D convolutional layers with $3 \times 3$ kernels. Inspired by \cite{demirel2020automatic}, we apply subsampling on the height axis after each alternate layer with a factor of 2, in order to get compact embeddings, $\mathbf{h}$, which are then fed into multiple streams of TDNNs\footnote{A time-delay neural network consists of 1-D convolutional layers where the convolution is applied with frames that are dilated on the time axis\cite{peddinti2015time}. In our architecture, we employ the factorized variant of TDNN introduced in \cite{povey2018semi}.}. Each stream of TDNNs has a unique time dilation rate, $\tau$, encoding information in different temporal resolutions,


 \begin{equation}
    \mathbf{z}^n_{t} = \mathit{Stacked{\text -}TDNN}(\mathbf{h} | \tau=t, N=n),
\end{equation}

\noindent where $\mathbf{z}^n_\tau$ are the latent variables at the output of the final ($N^\mathit{th}$) TDNN layer, and $t \in \mathbb{Z} $.  These are concatenated and projected to the classification ($\mathit{softmax}$) layer by a pair of fully connected ($\mathit{FC}$) layers,
 

\begin{equation}
    a(s) = \mathit{softmax}(2 \times \mathit{FC(Concat}(\mathbf{z}^N_{\tau_1},\mathbf{z}^N_{\tau_2},...,\mathbf{z}^N_{\tau_K}))),
\end{equation}

\noindent where $a(s)$ is the activation of the $\mathit{softmax}$ layer corresponding to the phoneme state $s$ and $K$ is the number of TDNN streams. We decide the number of layers per stream w.r.t.\ to the receptive field ($\mathit{RF}$) of the nodes at the top TDNN layer,


 \begin{equation}
    \mathit{RF}_{\mathbf{z}^N_\tau} = 2 \times l \times \tau \times N,
\end{equation}

\noindent where $l$ is the frame length of the acoustic feature vectors. Note that we include an additional 1-D convolutional layer just before the TDNN streams. This layer does not use dilation, in order not to skip any frames.

\section{Data}\label{sec:typeset_text}

\subsection{Training Set}\label{data:train}

The acoustic models of the previously presented ALT systems in the literature are built on either monophonic or polyphonic music recordings. In general, monophonic models are trained on the $\textit{DAMP}^\mathit{train}$ dataset\cite{dabike2019automatic,demirel2020automatic,demirel2021pron}, while $\textit{DALI}$ is utilized for polyphonic models\cite{gupta2020automatic}. We merge these two datasets, exploiting their size and musical variability.  We curated the polyphonic subset of the dataset on the recordings from the most recent version (v2.0) of the $\textit{DALI}$ dataset \cite{meseguerdata}, and included only those songs for which the Youtube links were available and still in use at the time of the audio retrieval.

\subsection{Evaluation Sets}

We perform model selection and optimization on the subsets of the $\textit{DAMP}$ and $\textit{DALI}$ datasets, representing monophonic and polyphonic domains respectively. For $\textit{DAMP}^\mathit{test}$, we use the test split introduced in \cite{dabike2019automatic}. For testing the lyrics transcription performance on polyphonic recordings, we curated a new subset of $\textit{DALI}$-v1.0, which we give the data selection procedure below. Finally, we evaluate our best performing model on the three benchmark datasets used in the literature, namely the Jamendo, Hansen and Mauch sets and provide a comparison with the state of the art in Section \ref{sec:sota}. 


\begin{table}[!h]
\renewcommand{\arraystretch}{1}
\centering
\rowcolors{2}{white}{gray!20}
\resizebox{0.47\textwidth}{!}{
\begin{tabular}{  c | c | c |  c | c | c | c | c}
  Set & Words & Uniq. Words & \# Utt. &  \# Rec   & \# Singers & Avg. Utt. Dur. & Total Dur.   \\
\hline
\textit{LM-corpus}  & 13M & 100k & 2M & N/A  & N/A  & N/A & N/A \\
\hline
 $\textit{DAMP}^\mathit{train}$  & 686k & 5.3k & 80k &  4.2k & 3k & 5sec & 112h \\
 $\textit{DALI}^\mathit{train}$  & 1.1M & 25.5k & 227k &   4.1k & 1.5k & 2.48sec & 156h \\
 \hline
 $\textit{DAMP}^\mathit{dev}$  & 4k & 695   & 482   & 66 & 38 & 5.12sec & 41min \\
 $\textit{DALI}^\mathit{dev}$  & 5.7k & 941 & 1.7k  & 34 & 16 & 2.41sec & 48min \\
\hline  
 $\textit{DAMP}^\mathit{test}$  & 4.6k & 840 & 479  & 70 & 40 & 6sec & 48min\\
  $\textit{DALI}^\mathit{test}$  & 62.8k & 4.2k & 240 & 240 & 160 & 233sec & 15.5h  \\
  \hline
   $\textit{Jamendo}$  & 5.7k & 1k & 20 & 20 & 20 & 216sec & 72min \\
   $\textit{Hansen}$  &  2.8k & 585 & 10 & 10 & 9 & 214sec & 35min \\
   $\textit{Mauch}$  & 5.2k & 820 & 20 & 20 & 18 & 245sec & 82min
   
\end{tabular}}
\caption{Statistics of datasets used in experiments }
\label{table:data}
\end{table}

For tuning the hyperparameters during evaluation, the language model scaling factor and the word insertion penalty, we have used the combination of the data from $\textit{DAMP}^\mathit{dev}$ split \cite{dabike2019automatic} and 20 recordings from $\textit{DALI}$-v2.0\footnote{This combined development set is denoted as \textit{dev} in Section \ref{sec:res}.}.

\subsubsection{The DALI-test set}

\begin{figure}[!h]
 \centering
 \includegraphics[clip,width=0.415\textwidth,height=2.22cm]{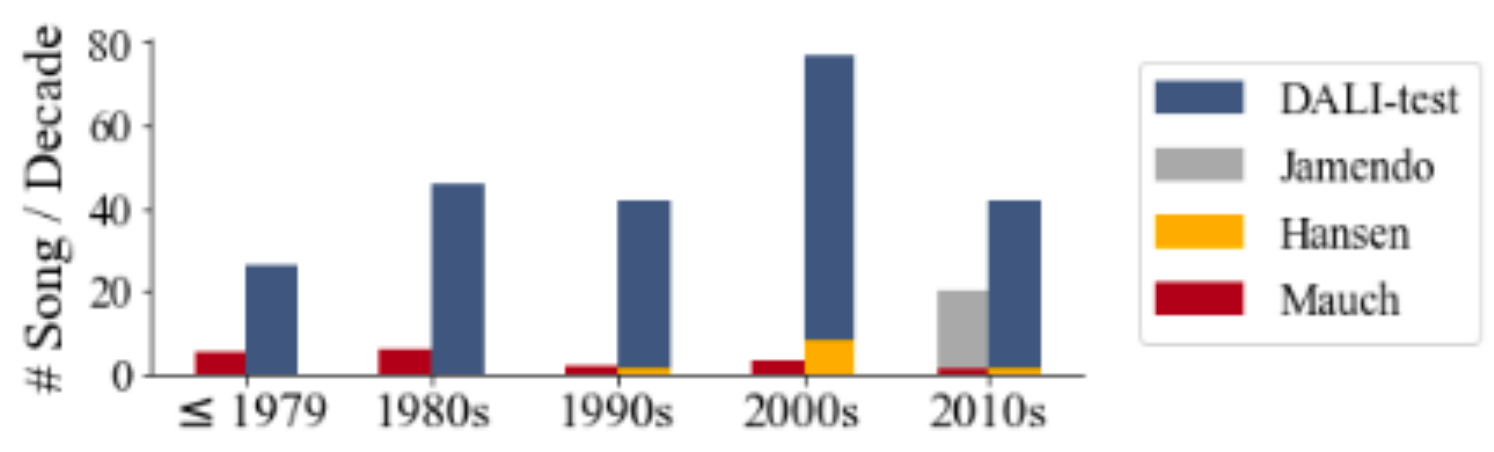}%
 \caption{Songs per decade in ALT evaluation sets}
 \label{fig:years}
\end{figure}

In this section, we give details of the curation procedure for the $\textit{DALI}^\mathit{test}$ set. We began from the subset presented in \cite{vaglio2020}, which initially had 513 recordings and filtered it according to a number of criteria. Numerous audio samples were not retrievable from the links provided. We obtained the Youtube links through \textit{automatic search} using relevant key words. We discarded songs where the automatically retrieved version was a live performance, had low audio quality or contained extra background speech sections unrelated to its corresponding lyrics. For consistency and fair evaluation, we did not include songs where the dominant language was not English. We allowed for an artist to have at most 5 songs. Among the remaining recordings, we manually selected a subset having a relatively balanced distribution of singers' gender, official release dates over decades (see Figure \ref{fig:years}) and variability in terms of singing styles, vocal effects and music genre. Lyrics were initially obtained from the annotations provided in \cite{meseguer2019dali} and manually verified following the steps explained in Section \ref{sec:lyr}. The final version of $\textit{DALI}^\mathit{test}$ consists of 240 recordings, which sets the largest test set for lyrics transcription with clean annotations. For open science, we publicly share the data identifiers, cleaned lyrics annotations and a tutorial to retrieve the corresponding Youtube links at ``\textit{https://github.com/emirdemirel/DALI-TestSet4ALT}''.


\section{Experimental Setup}

\subsection{Lyrics Preprocessing}\label{sec:lyr}

Prior to being utilized for training, raw  lyrics data automatically retrieved from online resources (as in $\textit{DALI}$) needs to be normalized, as the transcription rules applied by lyrics providers are not standardized. We remove all special ASCII characters except apostrophes. We convert numeric characters to their alphabetic correspondence. All text is converted to upper case. We observed several samples with erroneous hyphenation, explicit syllabification and repeating letters (possibly indicating longer uttered syllables or vowels). To cope with these, we apply automatic hyphenation correction and canonicalization using the standard open-source \textit{NLTK} tools\footnote{These steps are potentially language specific.}. The output lyrics are then verified and corrected manually.

\subsection{Language and Pronunciation Models}\label{sec:lm}

Lyrics often contain uncommon words that are not very likely to exist in standard pronunciation dictionaries. For such words that are not in the lexicon, or out-of-vocabulary (OOV) words, we generate pronunciations using a pretrained grapheme-to-phoneme (G2P) converter\cite{novak2012phonetisaurus}.  In order to produce fair and reproducible results, we generate pronunciations for the OOV words in the evaluation sets as well, and do not skip these during evaluation. We utilize the commonly used CMU English Pronunciation Dictionary\footnote{Link:  \textit{https://github.com/Alexir/CMUdict/blob/master/cmudict-0.7b}.} as the lexicon and generate alternative pronunciations by duplicating the vowel phonemes for each word pronunciation, inspired by the improvements observed in \cite{gupta2018automatic,demirel2021pron}. A 4-gram language model (LM) is constructed using the SRILM Toolkit \cite{alumae2010efficient}. We use the combination of the lyrics corpus in \cite{dabike2019automatic}\footnote{This corpus contains lyrics from all the songs of the artists included in the Billboard charts between 2015-2018, plus the lyrics in $\textit{DAMP}^\mathit{train}$.} and $\textit{DALI}^\mathit{train}$. For scientific evaluation, we exclude any songs which overlap with those in the evaluation sets.

\vspace{-0.15cm}

\subsection{Discriminating Instrumental and Silent Regions}

The hybrid DNN-HMM ASR framework approaches the continuous word recognition task essentially as a sequence decoding problem. Within this scope, the presence of instrumental accompaniment, especially during non-vocal regions, may disrupt the decoding path, potentially causing cumulative errors during transcription and alignment. Traditionally, non-speech regions are represented with a special silence token within the target class set during recognition. Here, we propose using separate tokens for the non-vocal instances in monophonic and polyphonic recordings. Prior to training, we associate these tokens with their corresponding silence/music instances by explicitly adding tags at the beginnings and the ends of the ground truth lyrics of each utterance (see Table \ref{table:tag}). These tags are represented as words in the lexicon where their pronunciations correspond to the relevant silence token.

\begin{table}[!h]
\renewcommand{\arraystretch}{1}
\centering
\resizebox{0.32\textwidth}{!}{
  \rowcolors{2}{white}{gray!25}
\begin{tabular}{ c | c  }
   \mc{2}{c}{\qquad $\mathbf{w}$}    \\
  Raw & $w_1$ $w_2$ ... $w_N$ \\
 $\textit{DAMP}$ & $<$\textit{silence}$>$ $w_1$ $w_2$ ... $w_N$ $<$\textit{silence}$>$ \\
 $\textit{DALI}$ & $<$\textit{music}$>$ $w_1$ $w_2$ ... $w_N$ $<$\textit{music}$>$
\end{tabular}}
\hfill
\caption{Music / silence tagging w.r.t.\ the dataset }
\label{table:tag}
\end{table}

For silence and music tagging, we exploit our training data. As we know the recordings in $\textit{DAMP}$ and $\textit{DALI}$ are monophonic and polyphonic respectively, we apply tagging w.r.t.\ the dataset. The pronunciations of these pseudo-word tags are represented with distinct phonemes in the lexicon.

\vspace{-0.15cm}


\subsection{Generating Phoneme Alignments}

Since the neural network optimization is performed w.r.t.\ phoneme posteriors (as explained in Section \ref{sec:lfmmi}), we need to extract their timings, i.e.\ \textit{alignments}. To generate these, we train a triphone Gaussian Mixture Model (GMM) - HMM  model on ``singer adaptive'' features \cite{anastasakos1996compact}, following the standard Kaldi recipe\footnote{We execute the GMM-HMM pipeline at \textit{https://github.com/emirdemirel/ALTA},
which is almost the same procedure as the standard \textit{librispeech} recipe, with tuned hyperparameters for singing data.}. At this stage, we compute per-word pronunciation probabilities following the steps in \cite{chen2015pronunciation}, and retrain another triphone model using the updated lexicon transducer. Using this new model, we apply \textit{forced alignment}\cite{gales2008application} on the training data for generating the phoneme and word alignments.

\subsection{Neural Network Training}

DNN training is 
based on the \textit{Kaldi - chain} recipe. In the feature space, we use 40-band filterbank features extracted with a hop size of 10ms and window size of 30ms. To achieve singer-adaptive training, we utilize i-Vectors\cite{saon2013speaker} which represent the singer identity information via global embedding vectors. Frame subsampling is applied with a factor of 3 in this training scheme where each subsampled frame in the input of the neural network is considered to represent $l = 3 \times 10$ms = $30$ms of context. The data is fed into the network as audio chunks of 4.2 seconds (140 frames) in minibatches of 32. We apply a decaying learning rate with beginning and final rates of $10^{-4}$ and $10^{-5}$ respectively. Stochastic gradient descent is used as the optimizer. The training is done for 6 epochs.

\vspace{-.26cm}

\section{Results}\label{sec:res}

We report lyrics transcription results based on word error rate (WER). We begin by comparing performances obtained using $\textit{DAMP}$ and/or $\textit{DALI}$ in training the acoustic model. Then we test our proposed idea of discriminating silence and accompaniment instances by using separate tokens, and perform experiments testing different topologies of the MTDNN architecture. To boost the performance further, we train a final model on augmented data and provide a comparison of our results with previously published models.

\vspace{-.16cm} 

\subsection{Multi-Domain Training}\label{sec:cd}

According to Table \ref{table:cd}, the model trained on $\textit{DAMP}^\mathit{train}$ performs relatively well on $\textit{DAMP}^\mathit{test}$, however its performance drops dramatically on polyphonic recordings. On the other hand, much better recognition rates are observed on $\textit{DALI}^\mathit{test}$ when a polyphonic model is used, but then the polyphonic model performs poorly on a cappella recordings. Finally, using recordings from both the monophonic and polyphonic domains results in improved performance on both polyphonic and monophonic test sets, although the improvement is marginal on the monophonic $\textit{DAMP}^\mathit{test}$ set.

\vspace{-.15cm}

\subsection{Music / Silence Modeling}

Next, we test whether the explicit music/silence tagging improves transcription results. At this stage, we use a single stream architecture ($M^\mathit{single}_8$ in Table \ref{res:nn}). Tagging is applied only in constructing the GMM-HMM model and for generating alignments. The music/silence tags were removed during neural network training. 
Table \ref{table:cd} shows that alignment with music/silence tags did result in considerably improved recognition results for polyphonic recordings, but no improvement was evident for the monophonic case. 

\begin{table}[!h]
\centering  \resizebox{0.3\textwidth}{!}{
\begin{tabular}{ c !{\vrule width 1.5pt } |c | c   }
 Train Set &  $\textit{DAMP}^\mathit{test}$ & $\textit{DALI}^\mathit{test}$ \\
 \hline
 $\textit{DAMP}$ & 17.64  & 78.42 \\\hline
 $\textit{DALI}$  & 61.95  & 59.19 \\\hline
 $\textit{DAMP}$ + $\textit{DALI}$ & \textbf{17.14}  & 53.86 \\
  \textbf{+} music/sil tag &   17.29  &  \textbf{47.00}  
\end{tabular}
 }
\caption{Multi-domain training and music/silence tagging results}
\label{table:cd}
\end{table} 


\subsection{Neural Architecture Design}

Here, we test various parameterizations of the multistreaming architecture. In this stage, we did not use the explicit music and silence tagging for training the models. As mentioned in Section \ref{sec:mtdnn}, we diversify each stream of TDNNs in terms of the number of hidden layers and/or their dimensions. In addition to achieving improved performance, the goal of these modifications is to exploit the temporal context to its full extent. For this, we calculate the number of TDNN layers included w.r.t.\ the resulting $\mathit{RF}_{\mathbf{z}^N_\tau}$.

In all MTDNN variants tested, we use 3 TDNN streams with $\tau \in \{3,6,9\}$. We begin with finding the optimal number of TDNN layers for the stream with the smallest $\tau$. For rapid experimentation, we use single-streaming TDNN models ($M^{single}_N$ in Table \ref{res:nn}). According to Table \ref{res:nn}, using 9 layers sets the optimal setup for $\tau=3$, having  $\mathit{RF}_{\mathbf{z}^{N=9}_{\tau=3}}=1620$ms. Note that further increasing the number of TDNN layers to 10 ($\mathit{RF}_{\mathbf{z}^{10}_{\tau=3}}=1800$ms) did not result in improved recognition, and the model complexity was much higher (Figure \ref{fig:res_perf}). Therefore, we chose as our baseline a single-stream model with 9 TDNN layers. 
\vspace{-0.3cm}

\begin{table}[!h], 
\scalebox{0.82}{
\centering 
\begin{adjustbox}{width=0.55\textwidth,height=2.45cm}
\small
\rowcolors{2}{white}{gray!25}
\def\arraystretch{1.5}
\begin{tabular}{ l l l  | l | l | l l }
& {Stream} &  {Layers}  & {Dimension} & {\textit{dev}}
& {$\textit{DAMP}^\mathit{test}$}  & {$\textit{DALI}^\mathit{test}$}   \\
 $M^\mathit{single}_7$  &   3    &  7  &    512  &  28.06 & 17.08  & 54.52 \\ 
 $M^\mathit{single}_8$ &   3    &  8  &    512  &  28.05 & 17.14 & 53.44 \\ 
 $M^\mathit{single}_9$ &   3    &  9  &    512  &  27.68 & \textbf{17.08} & \textbf{52.25} \\ 
 $M^\mathit{single}_{10}$  &   3    &  10  &    512  &  \textit{27.67} & 17.21 & 53.58 \\ 
 \hline
 $M^\mathit{multi}_{9,a}$ &  3-6-9   & (9,9,9) & (512,512,512) &  26.69 & 16.75 & 51.38 \\ 
 $M^\mathit{multi}_{9,b}$ &  3-6-9   & (9,4,3) & (512,512,512) &  \textit{26.65} & \textbf{16.45} & \textbf{49.32} \\ 
 $M^\mathit{multi}_{9,c}$ &  3-6-9   & (9,9,9) & (512,256,172) &  27.13 & 16.08 & 52.54 \\  
 $M^\mathit{multi}_{9,d}$ &  3-6-9   & (9,4,3) & (512,256,172) &  27.38 & 16.62 & 51.92 \\ 
\end{tabular}
  \end{adjustbox}
}
\caption{Experiments on NN design}
\label{res:nn}
\end{table}

\vspace{-0.3cm}

Next, we perform ablative tests on four variants of the multistream architecture (notated as $M^\mathit{multi}_{9,\{a,b,c,d\}}$). Model $M^\mathit{multi}_{9,a}$  have identical TDNN structures (except for $\tau$), whereas the variants $M^\mathit{multi}_{9,\{b,c\}}$ have reduced $N$ or hidden dimensions respectively w.r.t.\ $\tau$ at each stream. Both dimensions of model reduction are applied on $M^\mathit{multi}_{9,d}$. In models $M^\mathit{multi}_{9,\{b,d\}}$, we reduced the number of layers, $N$ for the streams with larger $\tau$ to keep $\mathit{RF}_{\mathbf{z}^N_{\tau}}$ similar across all streams. $M^\mathit{multi}_{9,\{b,d\}}$ have 4 and 3 layers at the streams with $\tau=6$ and $\tau=9$ having $\mathit{RF}$ values of 1440 and 1620ms respectively. On the other hand, adding one more layer on the streams with $\tau=6,9$ would result in having $\mathit{RF_{\mathbf{z}^{N}_{\tau}}} \geq $ 1800ms which is shown above to be suboptimal in the single-stream case (see results for $M^\mathit{single}_{10}$). 


\subsection{Model Selection}

   The proposed multistreaming setups except $M^\mathit{multi}_{9,c}$ outperformed their single-stream counterpart, $M^\mathit{single}_9$, particularly on $\textit{DALI}^\mathit{test}$. The best results are achieved with $M^\mathit{multi}_{9,b}$ which has unique $N$ layers across all streams with the same hidden layer dimension.

\begin{figure}[!h]
 \centering
 \includegraphics[clip,width=0.48\textwidth,height=2.9cm]{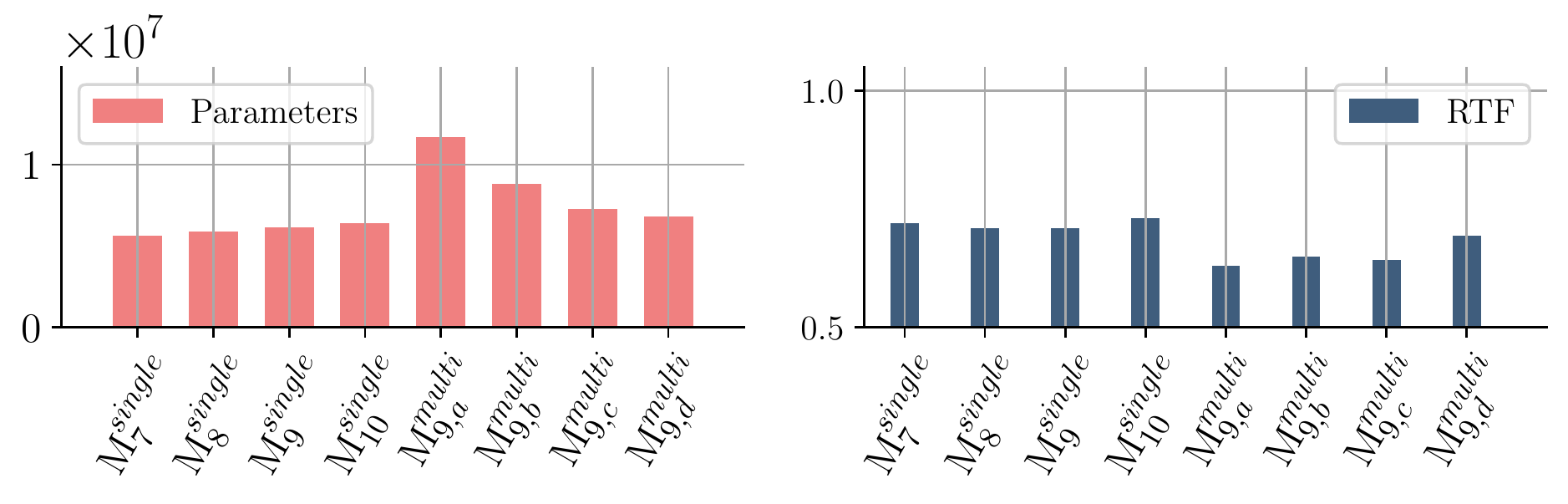}%
 \caption{Num. trainable parameters (left) \& RTFs (right)}
 \label{fig:res_perf}
\end{figure}

 To increase confidence in model selection, we investigate other operational aspects of the tested models. In Figure \ref{fig:res_perf}, we compare the number of trainable parameters which is a variable related to model complexity, and the real-time factor (RTF) that measures how fast a model operates during inference. We compute RTF's based on the inference times across all the data used in evaluation. We repeat this 5 times and report the mean of all iterations per model. These iterations are performed on an Intel$^{\text{\tiny{\textregistered}}}$ Xeon$^{\text{\tiny{\textregistered}}}$ 
 Gold 5218R CPU.
 


According to Figure \ref{fig:res_perf}, our best performing model $M^\mathit{multi}_{9,b}$ has the second largest number of trainable parameters. Its model complexity is however much lower than that of $M^\mathit{multi}_{9,a}$, the architecture presented in \cite{han2020multistream}. In terms of run time, all multistreaming models performed faster than single-stream models, with $M^\mathit{multi}_{9,b}$ being among the fastest. This shows that our compact variant has a reduced inference time with an improved recognition rate as hypothesized in Section \ref{sec:introduction}.

\subsection{Comparison with the State of the Art}\label{sec:sota}

At this last step, we train a final model combining the music / silence aware alignment with the best performing MTDNN architecture, $M^\mathit{multi}_{9,b}$. To boost the performance further, we apply data augmentation via speed perturbation with the factors of 0.9 and 1.1, tripling the size of the training data. In Table \ref{table:sota}, we compare our final model with other lyrics transcribers reported in the literature. We retrained the acoustic models in \cite{dabike2019automatic} ($M_{\text{\cite{dabike2019automatic}}}$), and \cite{demirel2020automatic} ($M_{\text{\cite{demirel2020automatic}}}$), using the corresponding publicly shared repositories. $M_{\text{\cite{gupta2020automatic}}}$ is based on the pretrained acoustic model shared at \textit{https://github.com/chitralekha18/AutoLyrixAlign}.  The same language model is used in constructing the decoding graphs for all the models in Table \ref{table:sota}. Note that $M_\text{\cite{dabike2019automatic},\cite{demirel2020automatic}}$ are trained on $\textit{DAMP}$ (monophonic) and $M_\text{\cite{gupta2020automatic}}$ is trained on $\textit{DALI}$ (polyphonic) datasets. We used the best performing language model scaling factor when reporting the results in Table \ref{table:sota}. We were not able to generate results on $\textit{DALI}^\mathit{test}$ using $M_\text{\cite{gupta2020automatic}}$ due to the model being highly memory intensive, as also reported in \cite{demirel2021low}.

In addition to these, we provide a comparison with the state of the art. The best WER score reported on $\textit{DAMP}^\mathit{test}$ is based on $M_{\text{\cite{demirel2020automatic}}}$ and applies rescoring on the word lattices generated after the first-pass decoding using an RNNLM\cite{xu2018pruned}. We did not apply RNNLM rescoring as we did not achieve consistent improvements across different test sets according to our empirical observations. For a fair comparison, we also include the best results in \cite{demirel2020automatic} achieved via n-gram LMs (the scores in paranthesis in Table \ref{table:sota}). On Jamendo, the best WER scores were reported in \cite{demirel2021low} where the inference was performed on source separated vocals. For Hansen and Mauch datasets, the best results are provided as reported in \cite{gupta2020automatic}\footnote{Note that the reason for the WER difference between $M_\text{\cite{gupta2020automatic}}$ and the scores reported in \cite{gupta2020automatic} is due to the bigger language model we used, despite both models having the same acoustic model.}. In order to evade optimistic results, we have discarded the overlapping songs between Hansen, Mauch and $\textit{DALI}^\mathit{train}$ during training the final model.


\vspace{-0.25cm}

\begin{table}[!h]
\centering
\scalebox{0.8}{
\setlength\tabcolsep{2.4pt} 
\begin{tabular}{ l l l | l l l  }
& \mc{5}{c}{\textit{WER}}  \\
\cmidrule{2-6}
& {$\textit{DAMP}^\mathit{test}$} & $\textit{DALI}^\mathit{test}$ & {Jamendo}  & Hansen & Mauch \\
\cmidrule{2-6} 
$M_{\text{\cite{dabike2019automatic}}}$ &  16.86   & 67.12 & 76.37 & 77.59 & 76.98 \\ 
$M_\text{\cite{gupta2020automatic}}$  & 56.90   &   N/A & 50.64 & 39.00 & 40.43 \\ 
$M_{\text{\cite{demirel2020automatic}}}$  &   17.16    &  76.72  & 66.96 & 78.53 & 78.50  \\ \cmidrule{1-6}
S.O.T.A & \textbf{14.96}  (17.01) \cite{demirel2020automatic} &  N/A & 51.76 \cite{demirel2021low}  & 47.01 \cite{gupta2020automatic} & 44.02 \cite{gupta2020automatic} \\
\cmidrule{1-6} 
MSTRE-Net & \textbf{15.38} &  \textbf{42.11} & \textbf{34.94}  & \textbf{36.78} & \textbf{37.33}
\end{tabular}}
\caption{Comparison with the state-of-the-art.}
 \label{table:sota}
\end{table}

\vspace{-0.3cm}

The results above show that MSTRE-Net outperforms all of the previously presented models on the polyphonic sets,  with more than 15\% , 7\% and 6\% absolute WER improvements achieved on the Jamendo, Hansen and Mauch datasets compared to the previous state of the art respectively. Notably, we achieved less than 50\% WER on the large $\textit{DALI}^\mathit{test}$ set indicating more than half of the words across 240 songs were correctly predicted. Our model also has the best results on $\textit{DAMP}^\mathit{test}$ achieved via n-gram LM. 

\vspace{-0.3cm}

\section{Conclusion}

We have introduced MSTRE-Net, a novel compact variant of the multistreaming neural network architecture, which outperforms previously proposed automatic lyrics transcription models. Our model achieved these results with lower model complexity and inference time. In addition, we showed that recognition rates improved across all evaluation sets after leveraging both polyphonic and monophonic data in training the acoustic model. We proposed a novel data preprocessing method for generating alignments prior to neural network training which resulted in considerably better word recognition rates from polyphonic recordings compared to the baseline approach. Finally, we curated a new evaluation set that is more comprehensive and varied, while having a much larger size compared to the previous test data used in research. For reproducibility and open science, the identifiers and a tutorial on making use of this data will be shared with the research community.

Our final model outperformed the previously reported best ALT results by a large margin, setting the new state-of-the-art. Through these results, we have taken an important step in increasing the potential and the possibility for ALT being an applicable technology in both Music Information Retrieval research and the music technology industry.

\end{document}